\documentclass[reqno,11pt,twoside]{article}  
\newcommand{\version}{January 2012}
\newlength{\dinwidth}
\newlength{\dinmargin}
\usepackage{amsmath}
\usepackage{amssymb}
\usepackage{latexsym}
\usepackage{cite}
\usepackage{epsfig,psfrag}
\newcommand{\RR}{\mathbb{R}}
\newcommand{\CC}{\mathbb{C}}

\newcommand{\calF}{{\mathcal F}}
\newcommand{\calG}{{\mathcal G}}
\newcommand{\calH}{{\mathcal H}}
\newcommand{\calO}{{\mathcal O}}

\newcommand{\calS}{{\mathcal S}}

\newcommand{\calR}{{\mathcal R}}
\newcommand{\Po}{{\mathcal P}_+^{\uparrow}}
\newcommand{\Potild}{{\mathcal P}^c} 

\newcommand{\He}{\calH^{(1)}}
\newcommand{\Ve}{V^{(1)}}

\newcommand{\half}{{\frac{1}{2}}} 
\renewcommand{\d}{{\rm d}}
\newcommand{\unity}{1}
\newcommand{\bfp}{{\boldsymbol{p}}}
\newcommand{\bfq}{{\boldsymbol{q}}}
\newcommand{\bfx}{{\boldsymbol{x}}}
\newcommand{\bfy}{{\boldsymbol{y}}}

\newcommand{\dom}{{\rm dom }} 

\newcommand{\supp}{{\rm supp}}
\newcommand{\eps}{\varepsilon}
\newcommand{\Om}{\Omega}
\newcommand{\OmN}{\Omega_0}
\newcommand{\lsp}{(\,}
\newcommand{\rsp}{\,)}

\renewcommand{\S}[1]{S(#1)}    
\newcommand{\SN}[1]{S_0(#1)}    
    
\newcommand{\FA}{\calF}     
\newcommand{\FAN}{\calF_0}  
 
\newcommand{\Ue}{U^{(1)}}

\newcommand{\HN}{\calH_0}  
\newcommand{\UN}{U_0} 
\newcommand{\Ee}{E^{(1)}} 
\newcommand{\spc}{C}   
   
\newcommand{\PFGs}{\calG}
\newcommand{\PFG}{G}
\newcommand{\GN}[1]{G_0(#1)}    
\newcommand{\FF}{\varphi_0}    
\newcommand{\Dom}{D}

\newcommand{\KTe}{K^{(1)}}     
\newcommand{\KTNe}{K_0^{(1)}}  
 
\newcommand{\V}{V}     
\newcommand{\spec}{\text{\rm sp}_P}     

\newcommand{\vN}{\calR}     
\newcommand{\In}{_{\text{\rm in}}}     
\newcommand{\Out}{_{\text{\rm out}}}     

\newenvironment{Proof}%
{\par \medskip \noindent {\em Proof.}}{\hspace*{\fill} $\square$\par%
\medskip\noindent}
\newtheorem{Thm}{Theorem}
\newtheorem{Prop}[Thm]{Proposition} 
\newtheorem{Lem}[Thm]{Lemma} 
\newtheorem{Cor}[Thm]{Corollary} 
\newtheorem{Ass}{Assumption}
\newtheorem{Remark}{Remark}

\newcommand{\beq}{\begin{equation}} 
\newcommand{\eeq}{\end{equation}} 
\newcommand{\pfg}{polarization-free generator }
\newcommand{\pfgs}{polarization-free generators }
\begin{document}
\title
{An Algebraic Jost-Schroer Theorem  for Massive Theories}
\author{Jens Mund\thanks{Supported by the Brazilian Research Council CNPq.}
\\ 
\scriptsize 
Departamento de F\'{\i}sica, Universidade Federal de Juiz de Fora,\\  
\scriptsize 
36036-900 Juiz de Fora, MG, Brazil.
}
\date{ \version \\ \vspace{3ex} 
{\em Dedicated to the memory of Claudio D'Antoni.}
}
\maketitle 
\begin{abstract}
We consider a purely massive local relativistic quantum theory
specified by a family of von Neumann algebras indexed by the 
space-time regions. 
We assume that, affiliated with the algebras associated to wedge regions, 
there are operators which create only single particle states from the
vacuum (so-called polarization-free generators) 
and are well-behaved under the space-time translations. 
Strengthening a result of Borchers, Buchholz and Schroer, we show that 
then the theory is unitarily equivalent to that of a free field for
the corresponding particle type. 
We admit particles with any spin and localization of the charge in 
space-like cones, thereby covering the case of string-localized 
covariant quantum fields. 
\end{abstract}
\maketitle 
\section*{Introduction} \label{secIntro} 
The Jost-Schroer theorem states that if the two-point function of a
Wightman quantum field is that of a free field, then it coincides with the
latter. 
The theorem has been proved by Schroer~\cite{S2}, Jost~\cite{Jo} and by
Federbush and Johnson~\cite{FederbushJohnson} for the massive case, 
and by Pohlmeyer~\cite{Po} for the massless case. 
Steinmann~\cite{St} has extended it to quantum fields which are
localized on strings (rays) with a fixed space-like direction as
explained below. 

Here, an analogous theorem is shown in the more general algebraic
framework, where the theory is specified by a family of von Neumann
algebras indexed by the space-time regions. 
Such an extension of the theorem is of relevance in view of the
existence of ``non-local'' models which do not correspond to point-localized 
Wightman fields, as realized in recent
years~\cite{BuSu_NonLoc,BuSu_Warped,BuLeSu_Warped,GrosseLechner07,
GrosseLechner08,MSY}.  
Another relevant aspect of the present article is that its results and
methods should be useful in the systematic construction of a quantum
field theory for Anyons with ``trivial'' S-matrix. 
Of course, both the hypothesis and the conclusion of the theorem need
appropriate modifications in the algebraic version. 
As to the hypothesis, we only consider theories which contain massive
particles (of arbitrary spin) separated by a gap from the rest of the
mass spectrum. 
The hypothesis of the Jost-Schroer theorem for Wightman fields is then
equivalent to the
condition that the fields generate from the vacuum only single particle states. 
So the appropriate hypo\-the\-sis in the algebraic setting is that there be  
operators which 

1.\ create only single particle states from the vacuum and 
are affiliated\footnote{A closed operator $G$ is said to be 
affiliated with a von Neumann algebra $\FA$ if all elements of the
commutant $\FA'$ of $\FA$ leave its domain invariant and commute with
$G$ on its domain.} with certain local algebras, and 

2.\ 
are well-behaved under the space-time translations in the
sense that they give rise to tempered distributions (see Section~\ref{secAss}).

In fact, we shall assume only the existence of such operators
affiliated with (Rindler) wedge regions\footnote{A wedge region is any
  Poincar\'e transform of the standard wedge $W_R$ which is
  characterized, in terms of a fixed Lorentz coordinate system, by 
\begin{equation} \label{eqW1} 
W_R \doteq \{\,x\in\RR^4:\,|x^0|<x^1\;\}. 
\end{equation}
}. 
Such operators have been called temperate polarization-free generators
(PFG's) 
by Borchers, Bucholz and Schroer~\cite{BBS}. 
These authors have shown that if there are temperate PFG's for wedge
regions, then the elastic two-particle scattering amplitude
vanishes in an open set, which in the case of compact charge 
localization 
implies that the S-matrix is trivial. It is well-known~\cite{BBS} that in 
any purely massive theory there are PFG's (satisfying only the
$1^{\rm st}$ property)  for any given wedge region,
in fact, there are sufficiently many as to create a dense set in the 
single-particle space. Thus our hypothesis only concerns the 
temperate behaviour of these operators under the translations 
($2^{\rm nd}$ property). 
Our conclusion is that then the net of local algebras is unitarily
equivalent to that of a free field for the corresponding particle 
type.\footnote{\label{JoSchAlg}It is an interesting open question
if the same conclusion also holds without
any temperateness assumption but if, instead, (non-temperate) PFG's are assumed 
to exist also in smaller regions, namely in space-like cones. 
}

We admit the most general localization properties for the charge
carried by the particles, namely, 
localization in space-like cones~\cite{BuF}.\footnote{A spacelike cone
  is a region in Minkowski space of the form 
 $\spc=a+\cup_{\lambda>0}\lambda \calO,$ where
  $a\in\RR^4$ is the apex of $\spc$ and $\calO$ is a double cone whose
  closure is causally separated from the origin.} 
(For simplicity, we consider only $U(1)$ inner symmetries.) 
Our result therefore also applies if the local field algebras are
generated by string-localized covariant quantum fields as envisaged 
in~\cite{St,MSY}, inspired by the ideas of 
Mandelstam~\cite{Mandelstam}: 
These are operator-valued distributions $\varphi_i(x,e)$ living on the
product of Minkowski space and the manifold of space-like directions $e$,
$e\cdot e=-1$. The fields are localized on strings $x+\RR_0^+e$
determined by the pairs $(x,e)$, namely, $\varphi_i(x,e)$ and $\varphi_j(x',e')$
commute if the corresponding strings are causally
separated. 
Steinmann has shown~\cite{St} the Jost-Schroer
theorem in the strict sense of Wightman fields for such fields, 
however only in the special case when the space-like directions of 
the strings all coincide, namely $e$ is fixed. 
In contrast, our result covers the case when the fields are 
actually distributions in $e$. 

The article is organized as follows. 
In Section~\ref{secAss}, we specify in more detail the general 
setting and the special assumptions, and present the result.  
In Section~\ref{secFF} we recall some facts on the free field algebras. 
The remaining two sections contain the proof of the Jost-Schroer
theorem: In Section~\ref{secCR} it is shown that the \pfgs
decompose into a creation and an annihilation part, 
and that the commutator of two \pfgs 
acts as a multiple of unity. 
This establishes the Fock space structure. 
The hermitean \pfgs turn out to be
self-adjoint, and hence unitarily equivalent to Segal field operators. The
remaining problem  (Section~\ref{secLoc}) is to identify the localized 
single particle vectors $\FF(f)\Omega_0$ of the free theory with
single particle vectors 
created by polarization-free generators. 
This is accomplished by a single particle version of the algebraic 
Bisognano-Wichmann theorem~\cite{M01a}. 
\section{Assumptions and Result}  \label{secAss} 
We start from a local relativistic quantum theory, specified by a
family $\spc\to \FA(\spc)$  of von Neumann algebras indexed by the 
space-time regions $\spc$ in a certain class. 
If all charges in the theory are strictly local, the class may be
taken to be the double cones, whereas we admit the presence of
topological charges~\cite{BuF}, hence the class will be taken to be 
  the space-like cones. The algebras $\FA(\spc)$  act in a Hilbert space
$\calH$ which carries a unitary representation $U$ of the universal 
covering group $\Potild$ of the Poincar\'e group with positive energy, {\it
  i.e.\ }the joint spectrum of the generators $P_\mu$ of the translations is
contained in the closed forward lightcone. There is a unique, up to a
factor,  invariant vacuum vector $\Omega$.   
The family $\spc\rightarrow\FA(\spc)$, together with the representation
$U$, satisfies the following properties. 
\\
$i)$ {\em Isotony:} $\spc_1\subset \spc_2$ implies 
 $\FA(\spc_1)\subset \FA(\spc_2).$  
\\
$ii)$ {\em Covariance:} For all  $\spc$ and all $g\in\Potild$ 
\begin{equation*} 
  U(g)\,\FA(\spc)\,U(g)^{-1}=\FA(g\,\spc) \,.  
\end{equation*}
(To simplify notation, we identify the action of the Poincar\'e 
group on Minkowski space with an action of its universal covering group.) 
\\
$iii)$ {\em Normal commutation relations:} 
There is a unitary ``Bose-Fermi'' operator $\kappa$, $\kappa^2=1$,  
leaving each field algebra $\FA(\spc)$ invariant, which 
determines the statistics character of field operators: 
Field operators which are even/odd under the adjoint action 
of $\kappa$ are Bosons/Fermions, respectively. 
Two field operators which are localized in
causally disjoint cones commute if one of the operators is bosonic and 
anti-commute if both of them are fermionic. 
This is equivalent to twisted locality~\cite{DHRI}: 
If $\spc_1$ and $\spc_2$ are spacelike separated, then 
\begin{equation} \label{eqTwiLoc}
 Z\FA(\spc_1)Z^* \subset \FA(\spc_2)'\,, 
\end{equation}
where $Z$ is the twist operator, $Z\doteq (1+i\kappa)/(1+i)$.
\\
$iv)$ {\em Reeh-Schlieder property:} For every $\spc$,  
 $\FA(\spc)\,\Omega$ is dense in $\calH$.  

We now specify our assumptions. 
As to the particle content of the theory, we assume that
there is one massive particle type (possibly with anti-particle) and
no massless ones. We thus make the 
\begin{Ass}[Massive particle spectrum.] \label{AssParticle}
The mass operator $\sqrt{P^2}$ has one isolated strictly positive 
eigenvalue $m$. The corresponding sub-representation $\Ue$ of $\Potild$ is
irreducible (neutral case) or has a two-fold degeneracy (charged case). 
\end{Ass}
We shall call the corresponding eigenspace of the mass operator the single
particle space and denote it by $\He$.
Our result easily extends to the more general situation of finitely
many particle types, and larger inner symmetry groups than $U(1)$. 
Note that by the spin-statistics theorem~\cite{BuEp}, the single
particle space must be 
fermionic (i.e., must be contained in the $(-1)$-eigenspace of $\kappa$) 
if the spin is half-integer, and bosonic otherwise.  
Our main assumption, namely the proper hypothesis of the
algebraic version of the Jost-Schroer theorem, concerns the
polarization-free generators. 
As mentioned in the introduction, it is well-known~\cite{BBS} that for 
every wedge $W$
there is a dense set of single particle vectors which are created from 
the vacuum by \pfgs localized in $W$. 
Namely, the dense set is the projection onto the single particle
space $\He$ of the domain of the Tomita operator $\S{W}$ associated with 
$\FA(W)$ and $\Omega$,\footnote{We recall the relevant notions on
  Tomita-Takesaki theory in the Appendix. 

The mentioned set is dense due to the commutation relations between 
the modular objects and the translations established by 
Borchers~\cite{Borchers92}.
}
and the \pfg $G$ which creates a given $\psi\in\dom \S{W}$ is the
closure of the operator 
\begin{equation} \label{eqGA'Om}
G_0\,A' \Omega = A' \psi, 
\quad A' \in \FA(W)'. 
\end{equation}
Following~\cite{BBS}, we 
call a \pfg $\PFG$ {\em temperate} if there is a dense subspace
$\Dom(G)$ of its domain, called its domain of temperateness, 
containing $\Omega$ which is invariant under the translations 
$U(x)\doteq U((x,\unity))$, and if for every $\psi\in\Dom(G)$ the function
$$
x\mapsto G\,U(x)\,\psi
$$ 
is continuous and polynomially bounded in norm
for large $x$, and the same holds for its adjoint $G^*$.  
Our main assumption now is that affiliated with every wedge algebra
$\FA(W)$ there are sufficiently many temperate \pfgs as to generate
a total subspace from the vacuum. 
In more detail, we assume the following. 
\begin{Ass}[Polarization-free generators.] \label{AssPFG} 
For each wedge region $W$ there is a self-adjoint\footnote{That means,
$\PFG\in\PFGs(W)$ implies $G^*\in\PFGs(W)$.} set $\PFGs(W)$
of temperate \pfgs affiliated with $\FA(W)$. 
Vectors of the form $\PFG_1\cdots \PFG_n\Omega$,
$\PFG_i\in\PFGs(W_i)$, are well-defined and contained in the domain of
temperateness of all polarization-free generators.\footnote{Actually,
  for our purpose it suffices to consider vectors of the above form
  where all $W_i$ contain some common space-like cone. 
}
For any fixed wedge $W$, the linear span of vectors of this form 
with $\PFG_i\in\PFGs(W)$ is dense in $\calH$. 
\end{Ass}
We shall denote the linear span of vectors of the form 
$\PFG_1\cdots \PFG_n\Omega$ with $G_i$ in some $\PFGs(W_i)$, or
respectively with all $G_i$ in the same $\PFGs(W)$, by 
$$
\Dom \; \text{ and }\; \Dom(W),
$$
respectively. 
%
Note that set $\PFGs(W)$ is invariant under the 
one-parameter family $U(\Lambda_W(t))$ representing the boosts 
$\Lambda_W(t)$ which leave the wedge invariant. For if $\PFG$ is in
$\PFGs(W)$, then for any $t$ the operator
$U(\Lambda_W(t))GU(\Lambda_W(-t))$ is also a \pfg affiliated with
$\FA(W)$, and it is also temperate due to the commutation relations of
boosts and translations~\cite{Borchers92}. 

The algebraic Jost-Schroer theorem which we are going to prove 
states that a theory $\FA$ satisfying our assumptions is unitarily 
equivalent to the net $\FAN$ of a free field for the corresponding particle 
type $(m,s)$. 
By a free field for the particle type at hand $(m,s)$ we mean
a Wightman field 
which generates from the vacuum only single particle 
states of the corresponding type, that is, which maps the
vacuum into the representation space $\He$ of the universal covering
group of the Poincar\'e group with mass $m$ and spin $s$. 
There are many such fields for a given particle type $(m,s)$, differing in the 
representation of $SL(2,\CC)$ according to which they transform. 
However, they all generate the same family of von Neumann algebras 
$W\to\FAN(W)$ associated with wedge regions up to unitary equivalence,
as we recall in Section~\ref{secFF}. 
We show: 
\begin{Thm}[Jost-Schroer theorem for wedge algebras] \label{JoSch}
Suppose the theory $\FA$ satisfies Assumptions~\ref{AssParticle} and
\ref{AssPFG}. Then there is a unitary operator $V$ from the Hilbert 
space $\calH$  onto the Fock space over $\He$,  
such that for any wedge $W$ and any element $g$ in the universal covering
group $\Potild$ of the Poincar\'e group there holds  
\begin{align} \label{eqFFW}
V\,\FA(W)\, V^*&=\FAN(W) , \\
V\, U(g)\,V^*      &=\UN(g), \label{eqUU0}\\
V\, \Omega &=\Omega_0.
\end{align}
Here, $\UN$ denotes the second quantization of the unitary
representation  $\Ue$ of $\Potild$ for mass $m$ and spin $s$ given by
the restriction of $U$ to $\He$, and $\Omega_0$
denotes the Fock space vacuum. 
\end{Thm}
Now the free field, and consequently the family $\FA$, both satisfy
twisted Haag duality for wedge regions~\eqref{eqTHDW}. Therefore 
their dual nets~\eqref{eqFd} are still local. The dual net $\FA^d$ is 
the maximal local extension of the net $\FA$ and has the same physical 
content as $\FA$. 
Since $\FA$ satisfies twisted Haag duality for wedges, the local dual
algebras are given by intersections over wedge algebras, 
$$
\FA^d(\calO)=\bigcap_{W\supset \calO}\FA(W),  
$$ 
see Remark~\ref{RemvN} (d). The same considerations hold for the free net. 
(In fact, $\FAN^d(\calO)$ coincides with $\FAN(\calO)$ at least for
neutral bosons, see Footnote~\ref{FFDual}.) 
Therefore we have 
\begin{Cor}[Equivalence of the local nets] 
The unitary $V$ from the theorem also implements, simultaneously for
all double cones $\calO$, the equivalence   
\begin{equation} \label{eqFF0}
V\,\FA^d(\calO)\, V^*=\FAN^d(\calO). 
\end{equation}
\end{Cor}
(In particular, of course, the algebras associated with double cones
are non-trivial.) 
\section{Remarks on the Free Field Nets}  \label{secFF}
A free field for mass $m$ and spin~$s$ is a Wightman field $\FF(f)$, 
$f\in\calS(\RR^4)\otimes\CC^N$, which acts on the (anti-) symmetrized Fock space
\begin{equation} \label{eqFockSpace} 
\HN \doteq  \Gamma_\pm(\He)
\end{equation}
over the single particle space $\He$ and transforms under some 
representation of $SL(2,\CC)$ acting on $\CC^N$. (The Fock space is
symmetrized or anti-symmetrized according to whether $s$ is integer or 
half-integer, respectively.)
The field operators $\FF(f)$, $f\in\calS(\RR^4)\otimes\CC^N$, are defined on a 
common dense domain $D_0$, the vectors with finite particle number. 
The representation of $SL(2,\CC)$ under which $\FF$ transforms and the property 
that the field creates from the
vacuum only single particle states with mass $m$ and spin $s$
characterize the corresponding free field up to unitary equivalence 
 --- this is the content of the Jost-Schroer theorem for Wightman fields.
We shall not need an explicit expression for the operators $\FF(f)$ in
any of these equivalent representations; what matters here is that 
they are necessarily of the form 
$\FF(f) =  a^*\big(\FF(f)\Omega_0\big)
+ a\big(\FF(f)^\dagger\Omega_0\big)$,\footnote{This must be so, again due to 
the Jost-Schroer theorem, since the two sides of this equation 
are Wightman fields with the same two-point function satisfying
the Klein-Gordon equation.}
where $\Omega_0$ is the Fock vacuum and $a^*(\phi)$ and $a(\phi)$,
$\phi\in\He$, denote the creation and annihilation operators in 
Fock space with domain $D_0$. (The dagger means $A^\dagger\doteq A^*|_{D_0}$.) 
Therefore the operator   
$\FF(f)+\FF(f)^\dagger$ 
coincides on $D_0$ with the Segal operator 
\begin{equation} \label{eqSegal}
\GN{\phi}\doteq  a^*(\phi)+a(\phi),    
\end{equation}
where $\phi = \big(\FF(f)+\FF(f)^\dagger\big)\Omega_0$, 
and is essentially self-adjoint on $D_0$~\cite{ReSi}. 
The fields $\FF(f)$ with $\supp f$ in a given space-time region
$\calO$ generate, in some sense, a von Neumann algebra
$\FAN(\calO)$. There are in principle various ways how this can be 
understood~\cite{BY90,BiWi,BaJuLl95}, but the {minimal requirement} is that the 
closures of $\FF(f)+\FF(f)^\dagger$ with $\supp f\subset \calO$ should be
{affiliated} with $\FAN(\calO)$. Hence the minimal choice is 
that $\FAN(\calO)$ be generated by these operators, namely
\begin{equation} \label{eqFNf}
\FAN(\calO)  =  
\big\{ e^{i\,\overline{\FF(f)+\FF(f)^\dagger}}|\; \supp f\subset \calO  \big\}''.  
\end{equation}
The corresponding net satisfies the Bisognano-Wichmann property and
hence twisted Haag duality~\eqref{eqTHDW} for wedge regions. 
{}For wedge regions $W$ the algebras $\FAN(W)$ can therefore not be chosen
any larger and \eqref{eqFNf} is the only possible 
choice.\footnote{\label{FFDual}The same holds for arbitrary contractible 
regions in the case of neutral free
bosonic fields, since in this case Haag duality has been 
shown~\cite{Araki64,LRT,Garber75} for the algebras~\eqref{eqFNf}. 
In the case of charged bosons~\cite{DellAntonio68} and
of fermions~\cite{DHRI,BaJuLl95,BaJuLl02}, duality has been shown for
a net which is defined in a different way than~\eqref{eqFNf} and it is not 
immediately clear whether the nets coincide. 
}

We now recall a characterization of $\FAN(W)$ in the context of 
Tomita-Takesaki theory.  
Let  $\SN{W}$ be the Tomita operator of $\FAN(W)$ and $\OmN$, 
and let  $\KTNe(W)$ be the intersection of its $+1$-eigenspace
with the single particle space. 
By construction, the vectors $(\FF(f)+\FF(f)^\dagger)\OmN$ 
with $\supp f$ in $W$ are contained in $\KTNe(W)$, hence 
$\FAN(W)$ is contained in the algebra generated
by the Segal operators $\GN{\phi}$ with $\phi\in \KTNe(W)$. 
But this algebra satisfies the Bisognano-Wichmann
property~\cite{BGL}. This has two consequences: Firstly, 
this apparently larger algebra in fact coincides\footnote{see  
Remark~\ref{RemvN}~(c).
}  with  $\FAN(W)$: 
\begin{equation} \label{eqFN}
\FAN(W) = \big\{ e^{i\GN{\phi}}|\; \phi\in \KTNe(W) \big\}'',
\end{equation}
and secondly,  $\KTNe(W)$ is fixed by the representation $\Ue$ of
$\Potild$ up to a unitary which commutes with
$\Ue$. Hence the algebra $\FAN(W)$ does not depend on the particular
free field which generates it, but only on the particle type, namely
on the representation $\Ue$.  
Note that the same argument shows that $\FAN(W)$ is generated,
in the sense of Eq.~\eqref{eqFN},  
by any real subspace of $\KTNe(W)$ which is invariant under the 
corresponding boosts and has  standard closure. 
\section{Fock Space Structure and Canonical (Anti-) Commutation Relations} 
\label{secCR}
By our Assumption~\ref{AssPFG}, for each $f\in\calS(\RR^4)$ and 
$\PFG\in\PFGs(W)$, where $W$ is an arbitrary wedge, there exists the 
strong integral 
\begin{equation*} 
\PFG(f)\doteq  \int d^4x f(x)\,\PFG(x),
\end{equation*}
where $\PFG(x)\doteq U(x)\PFG U(x)^{-1}$. 
The fact that $\PFG\Omega\in\He$ implies~\cite{M,BBS} that $\PFG(x)$
is a weak solution of the Klein-Gordon equation $\PFG((\Box+m^2)f)=0$
on the domain $\Dom$.  
This gives rise to a unique decomposition,
\begin{equation} \label{eqG+-}
G(f)=G^+(f)+G^-(f),
\end{equation}
namely the inverse Fourier transforms\footnote{The Fourier transform 
$\hat{f}$ of a test function $f$ is given by 
$\hat{f}(p) \doteq \int \d^4 x f(x) e^{ip\cdot x}$,  
and the inverse Fourier transform of the distribution $G$ is given
by $\check{G}(\hat{f}) \doteq G({f})$.
}
 of the distributions $G^+(\cdot)$ and 
$G^-(\cdot)$ are supported on the positive and negative  mass 
shells $H_m^{\pm}$ respectively.
By standard arguments~\cite{Jost}, this implies that for any test 
function $f$ the
operators $G^+(f)$ and $G^-(f)$ have momentum transfer in 
$H_m^+$ and $H_m^-$, respectively. 
In particular, therefore $G^-(f)$ annihilates the vacuum and maps
single particle vectors to multiples of the vacuum vector: If
$\phi\in \He$ is in the domain of temperateness of $G$, then 
\begin{equation} \label{eqG-1}
G^-(f)\phi = \lsp \Omega,G^-(f)\phi\rsp \,\Omega. 
\end{equation}
The action of $G^+(f)$ on single particle vectors has been largely
determined by Borchers, Buchholz and Schroer in~\cite{BBS}, namely, it
maps appropriate single particle vectors onto two-particle scattering
vectors. We state their result in a coordinate-free form. 
The given wedge $W$ defines an order relation\footnote{That is here a transitive
binary relation.} $\succ_W$ on 
the forward light cone 
$$
V_+ \doteq \{ p\in \RR^4\;\mid \, p\cdot p > 0,\; p_0>0\},  
$$
as follows. Let $a$ be some point in the edge, $E$, of $W$. Then $E-a$
is a space-like plane which contains the origin. For a given $q\in
V_+$, the set $(E-a) +\RR q$ is a 
time-like hyperplane which divides the forward light cone into two connected
components. The relation $p\succ_W q$ by definition  discriminates the component
which extends to space-like infinity in the same direction as $W$. In formulas: 
For $p,q\in V_+$ we write $p\succ_W q$ if and only if  
\begin{equation} \label{eqOrder}
p \,\in\,  (W-a) +\RR^+ q.
\end{equation}
Given compact sets $\V_1,\V_2\subset V_+$, we write $\V_1\succ_W\V_2$ if
$p\succ_W q$ for every $p\in\V_1$ and $q\in\V_2$. 
Now Borchers et al.~\cite{BBS} show that the domain of temperateness
of any \pfg contains single particle vectors
$\phi$ with arbitrarily small compact spectral support $\spec \phi$, 
and that for such vectors there holds:  
\begin{Lem}[\cite{BBS}] \label{BBS}
Suppose that the Fourier transform $\hat f$ of $f$ has support in a 
neighbourhood of a point on the positive mass shell $H_m^+$ 
(small enough as to contain no other spectral points). Then 
\begin{equation} 
G^+(f)\phi = 
\begin{cases} 
(G(f)\Omega\times\phi)\Out&\text{ if } 
\supp\hat f\succ_W \spec \phi\\
(G(f)\Omega\times\phi)\In& \text{ if }
  \spec \phi\succ_W \supp \hat f .
\end{cases}
\label{eqG+1Scatt} 
\end{equation}
\end{Lem} 
(We show in Appendix~\ref{secBBS} that this is equivalent with Lemma~3.2
in~\cite{BBS}.)  
Moreover, it turns out that the ``in''- and ``out''-states in 
Eq.~\eqref{eqG+1Scatt} coincide: 
\begin{Lem}[Triviality of the 2-particle S-matrix~\cite{BBS}] \label{STrivial}
The incoming and outgoing scat\-tering states constructed from any two
single particle states $\psi_1,\psi_2\in\He$ coincide: 
\beq \label{eqIn=Out}
(\psi_1\times \psi_2)\In =  (\psi_1\times \psi_2)\Out.
\eeq
\end{Lem}
This can be concluded from the work of Borchers et
al.~\cite{BBS}. They consider an incoming two-particle  
state with arbitrary momenta\footnote{Actually they consider $q_1$, $q_2$ in the
  center-of-mass-system of the form $(\omega(\bfq),\bfq)$,
  $(\omega(\bfq),-\bfq)$, with some condition on the relation of 
  $\bfq$ with the 1-axes. In view of covariance and our assumption of
  PFG's for all wedges, these restrictions are obsolete in the present context. 
} 
$q_1\neq q_2$, and show that there are outgoing momenta $p_1,p_2$ 
with $p_1+p_2=q_1+q_2$ such that 
${}\In\langle q_1,q_2|p_1,p_2\rangle\Out = 
{}\In\langle  q_1,q_2|p_1,p_2\rangle\In$. (In fact, both sides vanish
since their $\{p_1,p_2\}$ are disjoint from $\{q_1,q_2\}$.) 
The same conclusion is shown to hold if $q_1,q_2,p_1,p_2$ vary over sufficiently
small open sets. 
In the case of compact localization, it is well-known that this 
implies the asserted triviality of the 2-particle S-matrix~\cite{Araki,BEG64}. 
The proof uses the LSZ relations and momentum space analyticity of
both the time-ordered products and of the 
``intrinsic wave functions'' $\Ee B\Omega$.  
In the case of localization in space-like cones, these analyticity
properties 
are weaker~\cite{BrosEpstein94}, but still sufficient to prove the asserted
triviality of the 2-particle S-matrix~\cite{BrosMund}.\footnote{I am 
indebted to Jacques Bros, who has explained to me the argument, as
well as the existence of a gap in the case of non-compact 
localization, and  how it should be closed.}     

In the following we consider two wedges $W$ and $\hat W$ whose
intersection contains some space-like cone $C$. 
This situation entails two geometric facts. Firstly, the intersection of their
causal complements $W'\cap\hat W'$ also contains some spacelike cone
$\hat C$, cf.~Lemma~\ref{WedgeIntersect}, hence the algebra 
$\FA(W)'\cap \FA(\hat W)'$ contains a subalgebra 
for which the vacuum is cyclic, namely $Z\FA(\hat C)Z^*$.
Secondly, there are open sets $V_1,V_2$ in momentum space 
such that both relations $V_1 \succ_W V_2$ and $V_1\succ_{\hat W} V_2$
hold, cf.~Lemma~\ref{V1>V2}.   
Let now $G_1\in\PFGs(W)$ and $G_2\in\PFGs(\hat W)$ be two \pfgs and
let $f_1,f_2$ be test functions whose Fourier transforms have support 
in $V_1$ and $V_2$, respectively. 
Then the Lemmas~\ref{BBS} and
\ref{STrivial} imply that  
\begin{equation} \label{eqG1G2Om}
G_1^+(f_1)G_2(f_2)\Omega = 
\eps\, G_2^+(f_2)G_1(f_1)\Omega . 
\end{equation}
Here, and in the following, $\eps=1$ in the case of Bosons and
$\eps=-1$ in the case of Fermions.  
(To wit, the l.h.s.\ of Eq.~\eqref{eqG1G2Om} is 
$(G_1(f_1)\Omega\times G_2(f_2)\Omega)\Out$ and 
coincides with $\eps (G_2(f_2)\Omega\times G_1(f_1)\Omega)\In$ by the
(anti-) symmetry of the scattering states and Lemma~\ref{STrivial},
which is the r.h.s.) 
So the difference of the two sides of Eq.~\eqref{eqG1G2Om} is the
Fourier transform of a distribution supported on $H_m^+\times H_m^+$ which
vanishes on an open set.  
We now establish an analyticity property of this Fourier transform 
(extending Lemma~3.4 in~\cite{BBS}) which 
implies that it vanishes altogether, that is to say, that 
Eq.~\eqref{eqG1G2Om} holds for all $f_1,f_2$. 

In the Fourier decomposition of $G_1\in\PFGs(W)$ we shall use a
Lorentz frame $\{e_0,\ldots, e_3\}$ adapted to the wedge $W$: 
Let $e_2,e_3$ be an orthogonal basis of the 
space-like vector space $E-a$, where $E$ is the edge of $W$ and $a\in
E$, and let $e_0,e_1$ be a 
pseudo-orthogonal basis of the time-like plane orthogonal to $E-a$
such that $\RR^+e_1$ is contained in $W-a$. 
We shall denote by $x^\mu$ 
the corresponding contra-variant components of $x$,  
and write $\bfx\doteq (x^1,x^2,x^3)$, $\bfp\doteq (p^1,p^2,p^3)\in
\RR^3$. 
Since the Fourier transform of the function $G_1(x)$ is supported on
the (positive and negative) mass shells, its Fourier decomposition can
be written as  
\begin{align} 
G_1(x) & \equiv  \int d^4p \;\check G_1(p) e^{ip\cdot x} \nonumber \\
& = \int \frac{d^3\bfp}{2\omega(\bfp)}\, \big\{ 
\Gamma_1^+(\bfp)e^{i\omega(\bfp)x^0}+\Gamma_1^-(\bfp)e^{-i\omega(\bfp)x^0}\big\}
e^{-i\bfp\cdot \bfx}, \label{eqGGcheck}
\end{align}
where $\Gamma_1^\pm$ are (operator valued) distributions on $\RR^3$
and $\omega(\bfp)\doteq (\bfp^2+m^2)^{\half}$. 
Moreover, $\Gamma_1^\pm(\bfp)$ has momentum transfer in the
positive/negative mass shell, respectively. 
In the Fourier decomposition of $G_2\in\PFGs(\hat W)$ we shall use
coordinates $\hat x^\mu$ w.r.t.\ a basis $\hat e_\mu$ adapted 
to $\hat W$ in the analogous way, yielding 
\begin{equation} 
G_2(y) = \int \frac{d^3\hat \bfq}{2\omega(\hat \bfq)}\, \big\{ 
\Gamma_2^+(\hat \bfq)e^{i\omega(\hat \bfq)\hat y^0}+
\Gamma_2^-(\hat \bfq)e^{-i\omega(\hat \bfq)\hat y^0}\big\}
e^{-i\hat \bfq\cdot \hat \bfy}.  \label{eqGhatGcheck}
\end{equation}
In the following we write $\bfp^\perp\doteq (p^2,p^3)$ and
$\hat \bfq^\perp\doteq (\hat q^2,\hat q^3)$.  
\begin{Lem}[Momentum space analyticity]
Let $W$ and $\hat W$ be wedges as above,  
let $\{PFG_1\in\PFGs(W)$ and $\PFG_2\in\PFGs(\hat W)$, 
and let $A\in\FA(W)'\cap \FA(\hat W)'$. 
For $\bfp^\perp$ and $\hat \bfq^\perp$ in any compact set $M\subset \RR^2$ 
the distribution $\lsp A\Omega,\Gamma_1^+(\bfp)\Gamma_2^+(\hat \bfq)\Omega\rsp$ 
is the boundary value of an analytic function in the variables $p^1,\hat q^1$, 
analytic for $p^1,\hat q^1\in(\RR-i\RR^+)\cap U_M$, where $U_M$ is a
complex neighbourhood of the real axis. 
\end{Lem}
That is to say, for any $f^\perp$, $g^\perp\in\calS(\RR^2)$ with
supports in $M$ there is a function 
$(p^1,\hat q^1)\mapsto \lsp A\Omega,\Gamma_1^+(p^1, 
f^\perp)\Gamma_2^+(\hat q^1,g^\perp)\Omega\rsp$, analytic in
the mentioned strip, such that for every 
$f,g\in \calS(\RR)$ there holds 
\begin{multline*}
\lsp A\Omega,\Gamma_1^+(f\otimes f^\perp)\Gamma_2^+(
g\otimes g^\perp)\Omega \rsp = \\ \lim_{\eps\to
  0^+}\,\int dp^1d\hat q^1 f(p^1)g(\hat q^1) 
\lsp A\Omega,\Gamma_1^+(p^1-i\eps,f^\perp)\Gamma_2^+(\hat
q^1-i\eps,g^\perp) 
\Omega\rsp. 
\end{multline*}
\begin{Proof}
We consider the commutator function 
$$
K(x,y)\doteq  
\lsp A\Omega,G_1(x)G_2(y) \Omega\rsp-  
\lsp G_2^*(y)G_1^*(x) \Omega,A^*\Omega\rsp.
$$
By the remark before the lemma, the function $G_1(x)G_2(y)\Omega$ 
can be written as 
\begin{multline}  \nonumber 
G_1(x)G_2(y)\Omega =
\nonumber \\
 \int
\frac{d^3\bfp}{2p_0}\frac{d^3\hat \bfq}{2\hat q_0} \,  \big\{ 
\phi_+(\bfp,\hat \bfq)
\,e^{ip_0x^0} 
+\phi_-(\bfp,\hat \bfq)e^{-ip_0x^0}\big\}
e^{i\hat q_0\hat y^0} e^{-i(\bfp\cdot \bfx+\hat \bfq\cdot \hat \bfy)} ,
\label{eqGGcheck'}
\end{multline}
where $p_0\doteq \omega(\bfp)$, $\hat q_0\doteq \omega(\hat \bfq)$,  
and $\phi_\pm$ are the vector-valued distributions 
$\phi_\pm(\bfp,\hat \bfq)=\Gamma_1^\pm(\bfp)\Gamma_2^+(\hat \bfq)\Omega$. 
Similarly, $G_2^*(y)G_1^*(x)\Omega$ can be represented as 
\begin{multline*}
G_2^*(y)G_1^*(x)\Omega  = \\
\int
\frac{d^3\hat \bfq}{2\hat q_0}\frac{d^3\bfp}{2p_0} \,  \big\{ 
\psi_+(\hat \bfq,\bfp)\,e^{i\hat q_0\hat y^0} 
+\psi_-(\hat \bfq,\bfp)e^{-i\hat q_0\hat y^0}\big\}
e^{ip_0x^0} e^{-i(\bfp\cdot \bfx+\hat \bfq\cdot \hat \bfy)}
\end{multline*}
for some vector-valued distributions $\psi_\pm$. 
Then the commutator function has the form 
\begin{multline*}
K(x,y) = \int
\frac{d^3\bfp}{2p_0}\frac{d^3\hat \bfq}{2\hat q_0}  
e^{-i(\bfp\cdot \bfx+\hat \bfq\cdot \hat \bfy)} \times\\
 \times \big\{K_+(\bfp,\hat \bfq)e^{i(p_0x^0+\hat q_0\hat y^0)} 
+K_-(\bfp,\hat \bfq)e^{-i(p_0x^0+\hat q_0\hat y^0)} 
+K_0(\bfp,\hat \bfq)e^{-i(p_0x^0-\hat q_0\hat y^0)}\big\}, 
\end{multline*}
where the distributions $K_\pm,K_0$ are given by 
\begin{align*} 
K_+(\bfp,\hat \bfq)&= \lsp A\Omega,\phi_+(\bfp,\hat \bfq) \rsp,\\
K_-(\bfp,\hat \bfq)&= -\,\lsp \psi_+(-\hat \bfq,-\bfp),A^*\Omega \rsp,\\
K_0(\bfp,\hat \bfq)&= \lsp A\Omega,\phi_-(\bfp,\hat \bfq)\rsp-
\,\lsp \psi_-(-\hat \bfq,-\bfp),A^*\Omega \rsp .
\end{align*}
Now the restriction of the commutator function to the the 
time-zero plane $x^0=0=\hat y^0$ vanishes if $x^1$ and $\hat y^1$ are positive, 
since then $A$ and $G_1(0,\bfx)G_2(0,\hat \bfy)$ are
localized in causally separated regions. The same holds for its 
partial derivative with respect to $x^0$. Therefore the distributions 
$$
\frac{1}{\omega(\bfp)\omega(\hat \bfq)}(K_++K_- +K_0) 
\;\text{ and }\; 
\frac{1}{\omega(\hat \bfq)}(K_+-K_- - K_0) 
$$ 
are boundary values (in the sense explained before the proof) of
analytic functions in the variables $p^1$, 
$\hat q^1$, analytic in the lower half plane. 
Furthermore, for $\bfp^\perp$ in a compact set $M$, the function 
$p^1\mapsto \omega(\bfp)$ is analytic in some neighbourhood $U_M$ of the
reals.  Hence the distribution 
$K_+\equiv \lsp
A\Omega,\Gamma_1^+(\bfp)\Gamma_2^+(\hat \bfq)\Omega\rsp$ 
has the analyticity property claimed in the lemma. 
\end{Proof}
By Eq.~\eqref{eqG1G2Om}, the vector 
\begin{multline*}
G_1^+(f_1)G_2(f_2)\Omega - \eps G_2^+(f_2)G_1(f_1)\Omega \equiv \\  
\int \frac{d^3\bfp}{2p_0}\frac{d^3\hat \bfq}{2\hat q_0}\, 
\hat f_1(p_0,\bfp)\hat{f}_2(\hat q_0,\hat \bfq)\,
\big\{ 
\Gamma_1^+(\bfp)\Gamma_2^+(\hat \bfq)\Omega-\eps  
\Gamma_2^+(\hat \bfq)\Gamma_1^+(\bfp)\Omega 
\big\}, 
\end{multline*}
$p_0\doteq \omega(\bfp)$, $\hat q_0\doteq \omega(\hat \bfq)$, 
vanishes if 
the supports of $\hat f_1$ and $\hat f_2$ are contained in the open
sets $V_1,V_2$. 
By the analyticity property established in the last lemma and the fact
that $\big(\FA(W)'\cap \FA(\hat W)'\big)\Omega$ is dense (as observed 
before Eq.~\eqref{eqG1G2Om}), it vanishes altogether.  
In other words, Eq.~\eqref{eqG1G2Om} holds for any
$f_i\in\calS(\RR^4)$, $G_1\in\PFGs(W)$ and $G_2\in\PFGs(\hat W)$. 
Together with Eq.~\eqref{eqG-1}, this yields 
\begin{equation} \label{eqG1221Om}
[G_1(f_1),G_2(f_2)]_\eps\,\Omega = c(f_1,f_2)\, \Omega, 
\end{equation}
where $[A,B]_\eps\doteq AB-\eps BA$  
and $c(f_1,f_2)$ is the scalar product of $\Omega$ with the left hand side. 
The same relation holds of course for the non-smeared PFGs, and 
extends to the subspace $\Dom$ by standard 
arguments~\cite[proof of Thm.~4-3]{SW}:
\begin{equation} \label{eqG1221}
[G_1,G_2]_\eps  = c\, \unity \quad \text{ on } \Dom,
\end{equation}
where $c$ is the vacuum expectation value of the left hand side. 
This is the ``first half'' of the Jost-Schroer theorem, and
has already been shown in~\cite{M} under the assumption that temperate
\pfgs exist not only for wedges but for space-like cones, 
and only for PFGs localized in mutually causally separated
space-like cones. (This seems not sufficient for the present purpose
where we need the vectors of the form $G_1\cdots G_n\Omega$ to span a core.)  
Now recall that $G_1^-(f)$ may be written as $G_1(f_-)$, where
$\widehat{f_-}$ results from $\hat f$ by multiplication with a smooth 
function which is one on the negative and zero on the positive mass
shell. Hence 
\begin{equation*} 
[G_1^-(f),G_2]_\eps = c(f)\, \unity
\end{equation*}
holds on $\Dom$, where $c(f)$ is the vacuum expectation value of
the left hand side.  
Let now $W_1,\ldots, W_n$ be wedges whose intersection
contains some common space-like cone, 
\begin{equation} \label{eqW1WnC}
\bigcap_{k=1,\ldots,n}W_k \supset \spc,
\end{equation}
and let $G_k\in\PFGs(W_k)$. 
Taking into account that $G_1^-(f)$ annihilates the vacuum due to its
negative momentum transfer, the above considerations yield 
\begin{equation} \label {eqG-1nOm}
G_1^-(f)G_2\cdots G_n\,\Omega = \sum_{k=2}^n \eps^{k-1} 
c_k(f) \,G_2\cdots \widehat{G_k}\cdots
G_n\Omega   
\end{equation}
for any test function $f$, where the hat means omission of the corresponding 
factor, and 
$c_k(f)$ is the vacuum expectation value of $[G_1^-(f),G_k]_\eps$. To
determine this value, note that  $\widehat{\overline{f_-}}$ coincides with  
$\hat{\bar{f}}$ on the positive mass shell, hence
$G_1^-(f)^*\Omega=\hat{\bar{f}}(P) G_1^*\Omega$ and therefore 
$$
c_k(f) = \lsp \hat{\bar{f}}(P) G_1^*\Omega, G_k\Omega\rsp. 
$$ 
This shows that the operator-valued distribution $G_1^-(\cdot)$ is, on the
sub-space generated by vectors of the form $G_2\cdots G_n\,\Omega$, 
in fact given by a continuous function $G_1^-(x)$ and
that $G_1^-\doteq  G_1^-(x=0)$ also satisfies Eq.~\eqref{eqG-1nOm} with
$c_k(f)$ replaced by $\lsp G_1^*\Omega,G_k\Omega\rsp$. 
As is well-known from the usual Jost-Schroer theorem, the results
obtained so far imply that 
$\calH$ is isomorphic to the (anti-) symmetrized Fock space over
$\He$, and that $G_1^+$ and $G_1^-$ act as creation and annihilation 
operators, respectively (namely, $G_1^+$ 
creates $G_1\Omega$ and  $G_1^-$ annihilates $G_1^*\Omega$.) 
To set the stage for the next section, we recall the detailed argument.  

Equation~\eqref{eqG-1nOm}, with $G_1^-(f)$ replaced by $G_1^-$,
implies the recursion relation  
\begin{align}
\lsp \Omega,\PFG_1\cdots \PFG_{n}\Omega\rsp &=  
\lsp \Omega, 
\PFG_1^-\PFG_2\cdots\PFG_{n} \Omega\rsp \nonumber \\
&= \sum_{k=2}^n \eps^{k} \lsp G_1^*\Omega,G_k\Omega\rsp
\,\lsp\Omega, G_2\cdots \widehat{G_k}\cdots G_n\Omega\rsp \nonumber \\
&\equiv \lsp \Omega_0,\GN{\phi_1}\cdots \GN{\phi_n}\Omega_0\rsp,  
 \qquad \phi_k\doteq  \PFG_k\Om, \label{eqG1nOmNorm}
\end{align}
where $\OmN$ is the Fock space vacuum and $\GN{\phi}$ is the Segal
operator~\eqref{eqSegal}.  
The last equation holds because the canonical (anti-) commutation
relations imply the same recursion relation for the Segal
operators. 
Now an isomorphism between $\calH$ and the (anti-) symmetrized Fock space
$\HN$ over $\He$ is set up as follows. 
For any wedge $W$, let $\PFGs(W)^h$ denote the \pfgs $\PFG\in\PFGs(W)$ 
which satisfy 
$$
\PFG\Omega = \PFG^* \Omega.
$$
These \pfgs are hermitean, but in general (i.e., without our 
Assumption~\ref{AssPFG}) need not be self-adjoint. 
In any case, there holds 
\begin{equation} \label{eqG*GS}
\PFG^* = \PFG  \quad \text{ on } \Dom,
\end{equation}
since $\PFG^*$ and $\PFG$ both create the same 
vector from the vacuum and are affiliated with $\FA(W)$ (see the proof
 of Thm.~4-3 in~\cite{SW} for the argument). 
For a given wedge $W$, define an operator $V_W$ from 
$\Dom(W)$ into $\HN$ as follows. For $\PFG_k\in\PFGs(W)^h$, $k=1,\ldots,n$, let  
\begin{equation} \label{eqV}
V_W\, \PFG_1\cdots\PFG_n\Om\doteq 
\GN{\phi_1}\cdots\GN{\phi_n}\OmN \quad \text{ where } \phi_k\doteq  \PFG_k\Om,
\end{equation}
$V_W\Omega\doteq \OmN$, and extend by linearity to $\Dom(W)$. 
Eq.~\eqref{eqG1nOmNorm} implies that 
$V_W$ is well-defined and isometric. Since $\Dom(W)$ is dense in 
$\calH$ by Assumption~\ref{AssPFG}, 
$V_W$ extends uniquely to an isometric isomorphism from $\calH$ onto 
$\HN$. Let now $\hat W$ be another wedge such that $W\cap \hat W$
contains some space-like cone, and let $\hat \PFG_1,\ldots,\hat
\PFG_m\in\PFGs(\hat W)$. Then Eq.~\eqref{eqG1nOmNorm}
implies\footnote{Note that $W_1=\cdots=W_n\equiv W$ and
  $W_{n+1}=\cdots= W_{n+m}\equiv \hat W$ satisfy the
  condition~\eqref{eqW1WnC} under which Eq.~\eqref{eqG1nOmNorm} holds.} that 
$$
\|V_W\PFG_1\cdots \PFG_n\Omega- V_{\hat W}\hat\PFG_1\cdots \PFG_m\Omega\| = 
\|\PFG_1\cdots \PFG_n\Omega- \hat\PFG_1\cdots \PFG_m\Omega\|. 
$$
Therefore the closures of $V_W$ and $V_{\hat W}$ coincide. 
By iteration of the argument, one sees that the same holds true for
any pair of wedges $W,\hat W$. Thus, the closure of $V_{\hat W}$, which we
shall denote by $V_0$, is independent of $\hat W$ and satisfies for any
wedge $W$ and any $\PFG\in\PFGs(W)^h$ the intertwiner relation 
\begin{equation} \label{eqVGG0}
V_0\,\PFG\, V_0^* = \GN{\phi},\qquad \phi = \PFG\Omega ,
\end{equation}
on the dense sub-space $V_0\Dom(W)$, i.e., on the span of vectors of the form 
$\GN{\phi_1}\cdots\GN{\phi_n}\OmN$, $\phi\in\PFGs(W)$. 
The restriction of the Segal operator $\GN{\phi}$ to this subspace is
hermitean, and all these vectors are analytic for 
it~\cite[proof of Thm.~X.41]{ReSi}.  
By Nelson's theorem~\cite[Thm.~X.39]{ReSi}, $\GN{\phi}$
is therefore essentially self-adjoint on $V_0\Dom(W)$. 
By Eq.~\eqref{eqVGG0}, the hermitean operator $\PFG$ is essentially 
self-adjoint on $\Dom(W)$, and the 
unitary equivalence~\eqref{eqVGG0} of course holds for the self-adjoint 
operators $\PFG$ and  $\GN{\phi}$. To summarize, we have shown: 
\begin{Prop}[PFG's are Segal operators] \label{GGN}
Let $\PFG\in\PFGs(W)$ with $\PFG^*\Omega=\PFG\Omega$, where $W$ 
is any wedge region. Then $\PFG$ is essentially self-adjoint on 
$\Dom(W)$, and the unitary equivalence~\eqref{eqVGG0}
holds for the self-adjoint operators $\PFG$ and  
$\GN{\phi}$, where $\phi\doteq\PFG\Omega$.    
\end{Prop}
\section{Identification of the Local Algebras} 
\label{secLoc}
We have now identified certain \pfgs $\PFG$ with Segal field
operators $\GN{\phi}$.  
However, nothing has been said about the relation of these with the
local free field operators $\FF(f)$.  
We shall clarify this relation for wedge regions and in the context of 
Tomita-Takesaki theory, whose relevant notions we recall in 
Appendix~\ref{secTT}.
The unitary equivalence $V_0$ established in Eq.~\eqref{eqVGG0}
implements, simultaneously  for all wedges $W$, a unitary equivalence
of the von Neumann algebras
\begin{equation} \label{eqFhat}
\hat{\FA}(W) \doteq    
\big\{ e^{i\PFG}|\; \PFG\in \PFGs(W)^h \big\}''     
\quad \subset\FA(W)
\end{equation}
generated by the self-adjoint operators $G\in\PFGs(W)^h$, and the algebras 
\begin{equation} \label{eqFNhat}
\hat \FAN(W)\doteq  \big\{ e^{i\GN{\phi}}|\; 
\phi\in \PFGs(W)^h\Omega \big\}''. 
\end{equation}
The relation of $\hat \FAN(W)$ with $\FAN(W)$ as defined in
Eq.~\eqref{eqFNf} is not immediately clear,
and of course we shall not be able to identify a given vector $\PFG\Omega$, 
$\PFG\in\PFGs(W)^h$, with some $(\FF(f)+\FF(f)^\dagger)\OmN$, $\supp
f\subset W$.\footnote{This is the crucial difference from the case
  when $\FA$ is generated by a Wightman field $\varphi(f)$ which
  has the same transformation law under the Poincar\'e group as the 
free field $\FF(f)$: In this case, the single particle
  vectors $\varphi(f)\Omega$ are essentially fixed by covariance and
  coincide with $\FF(f)\Omega_0$ up to a unitary~\cite{Weinberg}. 
}
But it suffices to exhibit an isomorphism between the 
corresponding real subspaces of the single particle space.   
As mentioned at the end of Section~\ref{secFF}, $\FAN(W)$ is generated
not only by the set of vectors $(\FF(f)+\FF(f)^\dagger)\OmN$ 
but equally well by any real subspace of $\KTNe(W)$ with
standard closure which is invariant under the boosts corresponding 
to $W$. So the crucial point is to identify $\PFGs(W)^h \Omega$ with such a
subspace of $\KTNe(W)$.

Since the operators in $\PFGs(W)^h$ are 
affiliated with $\FA(W)$, the real space $\PFGs(W)^h\Omega$ is
obviously contained in the $+1$-eigenspace  
$$ 
\KTe(W)
$$ 
in $\He$ of the Tomita operator of $\FA(W)$. 
Far less obvious is the fact that $\KTe(W)$ and $\KTNe(W)$ essentially
coincide. At this point the algebraic version of the
Bisognano-Wichmann theorem~\cite{M01a} comes in: 
The family of algebras $\spc\to \FA(\spc)$ satisfies the 
hypothesis of~\cite[Prop.~3]{M01a} and therefore enjoys the 
Bisognano-Wichmann property on the single
particle space. In particular, the 
modular unitary group of $\FA(W)$ coincides on $\He$ with the representers of 
the boosts corresponding to the wedge $W$, and the modular conjugation 
represents the reflection at the edge of $W$, see Eq.~\eqref{eqJgJ}.  
Such operator is unique up to a 
unitary which commutes with all $\Ue(g)$, $g\in\Potild$. 
(Of course, in the neutral case it is a multiple of unity, and in the
charged case it is a unitary of the degeneracy space $\CC^2$.)
Now the Bisognano-Wichmann property is shared by the free field, 
and therefore the respective Tomita 
operators differ by a unitary which commutes with all $\Ue(g)$. 
We therefore have:
\begin{Prop}[Equivalence of the localization structures in
    $\boldsymbol{\He}$~\cite{M01a}]  
There is a unitary operator $\Ve$ which commutes with the
representation $\Ue$  of the universal covering group $\Potild$  of
the Poincar\'e group such that for all wedges $W$ there holds
$$
\Ve \KTe(W) = \KTNe(W).
$$
\end{Prop} 
Due to this result the real space $\Ve\PFG(W)^h\Omega$ is in $\KTNe(W)$. It
has standard closure and is
invariant under the boosts corresponding to $W$ since $\PFGs(W)$ is
invariant under these boosts and $\Ve$ commutes with them. Therefore,
as mentioned above, the algebra generated
by $\Ve\PFG(W)^h\Omega$ coincides with $\FAN(W)$. Summing up, the unitary
$V\doteq  \Gamma(\Ve)\circ V_0$ implements the equivalence 
$$
V\hat \FA(W)V^* = \FAN(W)
$$
simultaneously for all wedges $W$. 
It remains to show that  $\hat\FA(W)$ is not only a subalgebra but actually
coincides with $\FA(W)$. To this end, note that 
the family of free wedge algebras $\FAN(W)$ satisfies
twisted Haag duality~\eqref{eqTHDW}, hence the same holds for the
family $\hat\FA(W)$. Then $W\mapsto \FA(W)$ is local extension of the 
Haag dual (and therefore maximal) family of algebras 
$W\mapsto \hat\FA(W)$, which implies that
$\hat\FA(W)=\FA(W)$. (To wit: 
$$
\FA(W)\subset Z^*\,\FA(W')'\,Z \subset Z^*\,\hat\FA(W')'\,Z = 
\hat\FA(W) \subset \FA(W),
$$
which implies $\FA(W) = \hat\FA(W)$.) We therefore have shown that $V$
implements the unitary equivalence
\begin{equation} \label{eqFAFANW} 
V\, \FA(W)\,V^{*} = \FAN(W)
\end{equation}
simultaneously for all wedges $W$. Furthermore, $V$ commutes with the
representation of the universal covering group of the Poincar\'e group,
and thus the proof of our theorem is complete. 

\paragraph{Acknowledgments.}
It is a pleasure to thank Jacques Bros for valuable discussions. 
Thanks also to an anonymous referee for stimulating a relaxation of 
Assumption 2, and for raising the question mentioned in
footnote~\ref{JoSchAlg}.  
Further, I gratefully acknowledge financial support by the Brazilian
Research Council CNPq.  
\appendix
\setcounter{Remark}{0}
\setcounter{Thm}{0}
\setcounter{equation}{0}
\renewcommand{\theequation}{\thesection.\arabic{equation}}
\renewcommand{\theRemark}{\thesection.\arabic{Remark}}
\renewcommand{\theThm}{\thesection.\arabic{Thm}}
\section{Tomita-Takesaki Theory and the \\ Bisognano-Wichmann Theorem} 
\label{secTT}
Let $\vN$ be a von Neumann algebra of operators acting in some Hilbert
space $\calH$, and let $\Omega\in\calH$ be a cyclic and separating vector for
$\vN$. The {\em Tomita operator} associated with $\vN$ and $\Omega$ is
the closed operator $S$ characterized by 
$$
S\, A\Omega =  A^*\Omega, \quad A\in\vN.
$$ 
This is an anti-linear, densely defined, closed operator which is 
involutive, {i.e.}, satisfies $S^2\subset \unity$. It is therefore
completely determined by its $+1$ eigenspace, 
$$
 K \doteq \{\,\phi\in \dom S:\,S\,\phi=\phi\,\} \;. 
$$
For every vector in the domain of $S$ may be uniquely written as
$\psi=\phi_1+i\phi_2$ with $\phi_1,\phi_2\in K$, 
namely $\phi_1\doteq\half(\psi+S\psi)$ and $\phi_2\doteq \frac{1}{2i}
(\psi-S\psi)$, and obviously $S(\phi_1+i\phi_2)=\phi_1-i\phi_2$ 
for $\phi_1,\phi_2\in K$. 
In particular, $K$ is a real closed subspace of $\calH$ which is {\em
  standard}, namely $K+iK$ is dense in $\calH$ and $K\cap iK=\{0\}$. 

Writing the polar decomposition of $S$ as $S=J\Delta^{1/2}$, the operator
$J$ is an anti-unitary involution called the modular
conjugation, and the positive operator $\Delta$ gives rise to the
so-called {\em modular unitary group} $\Delta^{it}$ associated with 
$\vN$ and $\Omega$. 
The Tomita-Takesaki theorem asserts that the adjoint action of the
modular unitary group leaves the algebra $\vN$ invariant, while
that of the modular conjugation maps $\vN$ onto its commutant. 
The following facts are relevant in our context:
\begin{Remark} \label{RemvN}
 \rm 
a) 
Let $\hat{\vN}$ be a von Neumann sub-algebra of $\vN$, for which
$\Omega$ is still cyclic and which is invariant under the adjoint
action of the modular unitary group of $\vN$ and $\Omega$. Then
$\hat{\vN}=\vN$. (For then $\hat\vN\Omega$ is dense and invariant
under $\Delta^{it}$ and therefore is a core for
$\Delta^{1/2}$~\cite[Thm.~VIII.11]{ReSi}, which implies that
$\hat\vN=\vN$~\cite[Thm.~9.2.36]{KadisonRingrose}.)    
  
b) 
The {\em Bisognano-Wichmann theorem}~\cite{BiWi,M01a} 
states that for a large class of models
$\{\calR(\spc)\}_\spc$ the modular objects $J$ and $\Delta$
associated with the algebra $\calR(W_R)$ of the standard wedge region 
$W_R$ and the vacuum $\Omega$ are
related to the representation of $\Potild$ under which the fields
transform as follows. 
The modular unitary group coincides with
the unitary one-parameter group representing the boosts
$\lambda_1(\cdot)$ which leave the wedge $W_R$ invariant, 
namely $\Delta^{it}=U(\lambda_1(\alpha
t))$ for some real constant $\alpha$. (We call this relation the {\em
  Bisognano-Wichmann property}.)  
Further, the anti-unitary operator $J$ represents the 
reflection $j=\text{ diag }(-1,-1,1,1)$ at the edge of the wedge in
the sense that $J^2=\unity$ and 
\begin{equation} \label{eqJgJ}
J U(g) J  = U(jgj)
\end{equation}
holds for all $g\in\Potild$. 
As a consequence of these relations, the field 
satisfies {\em twisted Haag duality} for wedge regions, i.e., 
\begin{equation} \label{eqTHDW}
Z\,\calR(W)\,Z^* =  \calR(W')'. 
\end{equation}
The Bisognano-Wichmann theorem holds for the free fields, and
has been shown to hold in the algebraic setting for any theory
satisfying our Assumption~\ref{AssParticle} and asymptotic 
completeness~\cite{M01a}. 

c)
Let $\{\calR(\spc)\}_\spc$ be a model which satisfies the
Bisognano-Wichmann property. Suppose that for some wedge $W$, 
$\calR_0(W)$ is a von Neumann subalgebra of
$\calR(W)$ for which the vacuum is still cyclic and which is invariant under the
representers of the boosts leaving $W$ invariant. Then $\calR_0(W)$
coincides with $\calR(W)$ by part (a) of the remark. 

d) Twisted Haag duality for wedges implies that the so-called dual
net,  defined by 
\begin{equation} \label{eqFd}
\vN^d(\calO)\doteq  \big(Z^*\vN(\calO')Z\big)',
\end{equation}
is still local and that it coincides
for convex causally complete regions $\calO$ with 
\beq \label{eqDualW}
\vN^d(\calO) = \bigcap_{W\supset\calO}\vN(W),
\eeq
where the intersection goes over all wedge regions containing
$\calO$. (The proof goes as that of Cor.~3.5 in~\cite{BGL}.)   
It is the maximal local extension of the net $\vN(\calO)$ in the sense
that it contains every local extension of it, and has the same
physical content, for example the same S-matrix. 
\end{Remark}
\section{Geometric Considerations} \label{secGeo}
\setcounter{Remark}{0}
\setcounter{Thm}{0}
\setcounter{equation}{0}
\subsection{Intersections of Wedges.} 
In section~\ref{secCR} we used some rather obvious properties of the class of
space-like cones and wedges which we prove here for completeness' sake.
\begin{Lem} \label{CW}
Let $\spc$ be a space-like cone with apex $b$, and let $W$ be a 
wedge with apex $a$. (That is, $a$ is contained in the edge of $W$.)  
Then $C\subset W$ if and only if $b\in \overline{W}$
and $C-b\subset W-a$. 
\end{Lem}
\begin{Proof}
Since both $C$ and $W$ are convex, $C\subset W$ is equivalent with
$\overline{C}\subset \overline{W}$. Now
$\overline{C}=b+\RR_0^+\overline{\calO}$ where $\overline{\calO}$ is
causally separated from the origin. Let
$e\in\overline{\calO}$. According to Lemma~A.1 of \cite{M01a}, 
$b+\RR_0^+e\subset {W}$ is equivalent with $b\in W$ and
$e\in\overline{W}-a$. With the same method one shows that 
$b+\RR_0^+e\subset \overline{W}$ if and only if 
$b\in \overline{W}$ and $e\in\overline{W}-a$. Hence
$\overline{C}\subset \overline{W}$ is equivalent with
$b\in\overline{W}$ and $\overline{\calO}\subset\overline{W}-a$.  
But $\overline{C}\subset \overline{W}\Leftrightarrow C\subset W$, and 
$\overline{\calO}\subset\overline{W}-a \Leftrightarrow \subset
  W-a\Leftrightarrow C-b\subset W-a$, which proves the claim. 
\end{Proof}
\begin{Lem} \label{WcapW}
Let $W_0$ and $\hat W_0$ be wedges whose edges contain the origin, and
let $C_0$ be a space-like cone with apex at the origin such that
$C_0\subset W_0\cap \hat W_0$. Then 
for every $a,\hat a\in\RR^4$ there exists some $b\in\RR^4$ such that
\begin{equation} \label{eqCbinWa}
C_0+b\,\subset (W_0+a)\cap (\hat W_0+\hat a). 
\end{equation}
\end{Lem}
\begin{Proof}
By Lemma~\ref{CW}, it is sufficient to show that the closures of 
$W_0+a$ and $\hat W_0+\hat a$ have non-empty intersection. 
As a first step, note that for every $e\in W_0$ and $\check a\in
\RR^4$ there holds $\lambda e\in W_0+\check a$ for all sufficiently
large $\lambda\in\RR$. 
(This can be verified taking $W_0$ the standard wedge. Then
$\lambda e\in W_0+\check a$ if and only if $\lambda e^1>\check a^1$ and $\big(
(e^1)^2-(e^0)^2\big) \lambda^2+ \beta \lambda+\gamma>0$, where $\beta$ and
$\gamma$ are some real numbers. Since $e^1> 0$ and
$(e^1)^2-(e^0)^2>0$ by hypothesis, both conditions are satisfied for 
sufficiently large $\lambda$.) 
This result implies that for every $e\in W_0\cap \hat W_0$ (which
is non-empty by hypothesis) there is some $\lambda\in\RR^+$ such that 
$$ 
\lambda e\in (W_0+a-\hat a)\cap \hat W_0.
$$
Then $b:=\lambda e +\hat a$ is contained in $(W_0+a)\cap
(\hat W_0 +\hat a)$, completing the proof.
\end{Proof}
\begin{Lem} \label{WedgeIntersect}
If the intersection of two wedges $W\cap \hat W$ contains some 
space-like cone $C$, then the intersection of their causal complements 
$W'\cap\hat W'$ contains the space-like cone $-C+c$ for some
appropriately chosen $c\in\RR^4$.
\end{Lem}
\begin{Proof}
By Lemma~\ref{CW}, the hypothesis implies that 
$(W-a)\cap (\hat W-\hat a)$ contains $C-b$, that is, 
$-C+b\subset (-W+a)\cap (-\hat W+\hat a)$. Here $a,\hat a$ are in the
respective edges of $W, \hat W$ and $b$ is the apex of $C$.  
Now the cone and wedges involved have apex and edges at the origin. 
Hence Lemma~\ref{WcapW} applies, to the effect that there is some 
$\hat b$ such that  
$-C+b+\hat b\subset (-W+2a)\cap (-\hat W+2\hat a)$. 
But $-W+2a$ is just $W'$, and similarly for $\hat W'$, and thus 
the proof is complete.
\end{Proof}
\subsection{The Order Relation on Velocity space.} 
Our definition~\eqref{eqOrder} of the order relation  $p\succ_W q$ 
induced by a wedge $W$
is obviously equivalent with 
\begin{equation} \label{eqOrder''}  
p \,\in\,  (E-a) + \RR^+e +\RR^+ q,
\end{equation}
where $E$ is the edge of $W$, $a\in E$, and where $e$ is any fixed
vector in the interior of $W-a$. 
In this form, one sees easily that the relation $\succ_W$ is transitive.
Further, relation~\eqref{eqOrder''} is obviously equivalent 
with $q \,\in\,  (E-a) - \RR^+e +\RR^+ p$, hence with 
\begin{equation} \label{eqq<p}
q \,\in\, - (W-a) +\RR^+ p.
\end{equation}
We therefore have 
\begin{Lem} \label{q<p}
The relation $\succ_W$ defined on $V_+$ by Eq.~\eqref{eqOrder} 
is transitive. The relation $p\succ_W q$  is equivalent with $q\succ_{-W} p$. 
\end{Lem}
In Section~\ref{secCR} we have used the following fact. 
\begin{Lem} \label{V1>V2}
Let $W$ and $\hat W$ be wedges whose intersection contains some
space-like cone $C$. Then there are open subsets
$V_1,V_2$ of $V_+$ which have non-trivial intersection with $H_m^+$ 
and satisfy both $V_1\succ_W V_2$ and $V_1\succ_{\hat W}V_2$. 
\end{Lem}
\begin{Proof}
Let $a$ and $\hat a$ be points in the edges of $W$ and $\hat W$
respectively. 
By Lemma~\ref{CW}, $(W-a)\cap (\hat W-\hat a)$ contains the cone 
$C-b=: C_0$, where $b$ is the apex of $C$. 
Fix some velocity vector $k\in V_+$, and define 
$$
V_1\doteq (C_0+\RR k)\,\cap
V_+, \quad  V_2\doteq (-C_0+\RR k)\,\cap V_+ . 
$$
Then $V_1\succ_W k$ and $V_1\succ_{\hat W}k$. Further, by
Lemma~\ref{q<p} or Eq.~\eqref{eqq<p}, $k\succ_{W}V_2$ and $k\succ_{\hat W} V_2$. 
By transitivity (Lemma~\ref{q<p}), this proves the claim.
\end{Proof}
For the proof that our order relation is the covariant generalization
of the order relation used in \cite{BBS}, we shall need: 
\begin{Lem} \label{p>q}
For the standard wedge $W_R$, the relation $p\succ_{W_R}q$ is equivalent with 
\begin{equation} \label{eqOrder1}
\frac{p^1}{p^0}>\frac{q^1}{q^0}.
\end{equation}
\end{Lem}
Here, $p^\mu$ denote the contra-variant components of $p$ with respect
to the Lorentz frame $\{e_0,\ldots,e_3\}$ adapted to $W_R$ (i.e., the
unit vectors $e_2$ and $e_3$ span the edge of $W_R$ and  the ray $\RR^+e_1$ is
contained in $W_R$). 
\begin{Proof}
We first show that 
\begin{equation} \label{eqOrderE}
\frac{p^1}{p^0}=\frac{q^1}{q^0}\; \Leftrightarrow \; p\in E + \RR q,
\end{equation}
where $E$ is the edge of $W_R$. 
The direction ``$\Leftarrow$'' is clear, and we show
``$\Rightarrow$''. 
Let $\bar e_0,\bar e_1$ be any pseudo-orthogonal basis of the orthogonal
complement $E^\perp$ such that $\bar e_0$ is time-like future pointing 
and $\bar e_1$ points into the same direction as $e_1$, 
that is, $\RR^+ \bar e_1\subset W_R$. 
By application of a boost one can transform the basis
$\{e_0, e_1\}$ of $E^\perp$ into $\{\bar e_0,\bar e_1\}$. 
But the condition $\frac{p^1}{p^0}=\frac{q^1}{q^0}$
is invariant under these boosts,
and hence equivalent with $\frac{\bar p^1}{\bar p^0}=\frac{\bar
  q^1}{\bar q^0}$, where $\bar p^\mu$ are the (contra-variant)
coordinates  of $p$ with respect to $\{\bar e_0,\bar e_1\}$.  
(For 
$$
\frac{(\Lambda_1(t)p)^1}{(\Lambda_1(t)p)^0}=
\frac{p^1+\tanh(t)p^0}{p^0+\tanh(t) p^1} 
$$
and $\frac{p^1+c p^0}{p^0+c p^1}=\frac{q^1+c q^0}{q^0+c q^1}$ if and
only if $\frac{p^1}{p^0}=\frac{q^1}{q^0}$.)    
Now let, specifically, $\bar e_0, \bar e_1$ be the pseudo-orthogonal 
basis of the orthogonal complement $E^\perp$ such that $\bar e_0$ is
the normalized projection of $q$ onto $E^\perp$. Then $q$ lies in the
linear span of $E$ and $\bar e_0$, hence 
$\bar q^1=0$. Then the condition 
$\frac{\bar p^1}{\bar p^0}=\frac{\bar q^1}{\bar q^0}$ 
implies that $p$ also lies in the linear span of $E$ and $\bar e_0$,
which coincides with $E+\RR q$. This proves the equivalence~\eqref{eqOrderE}. 

Now $E +\RR q$ is a time-like hyperplane which divides $V_+$ into two 
connected components. 
The relation $\frac{p^1}{p^0}>\frac{q^1}{q^0}$ 
discriminates (for fixed $q$) those $p$ contained the component which 
extends to space-like infinity 
in the direction $e_1$. But the same component is discriminated by the
relation $p\succ_{W_R} q$. Therefore the relations coincide. 
\end{Proof}
\subsection{Proof of Lemma~\ref{BBS}.}  \label{secBBS}
We now show that our Lemma~\ref{BBS} is equivalent with Lemma 3.2\ of
Borchers et al.~\cite{BBS}. 

Suppose that the Fourier transform $\hat f$ of $f$ has support in a 
neighbourhood of a point on the positive mass shell $H_m^+$ 
(small enough as to contain
no other spectral points). Borchers et al.~\cite{BBS} 
define the velocity support $\V(f)$ of
$f$, with respect to some fixed time-like unit vector as reference frame, by  
\begin{equation}  \label{eqVf}
\V(f) \doteq  \{\, \big(1,\frac{\bfp}{\omega(\bfp)}\big)\,,\;
p=(p_0,\bfp)\in\supp\hat{f}\,\}\, ,   
\end{equation}
where $\omega(\bfp)\doteq (\bfp^2+m^2)^{\half}$. The velocity
support $\V(\phi)$ of a single particle vector $\phi\in\He$ is defined
analogously, with $\supp\hat f$ replaced by the spectral support
$\spec\phi$ of $\phi$. 
Borchers et al.~\cite{BBS} consider the standard wedge $W_R$ and
define a partial order relation $>$ on
velocity space, namely, given compact sets $\V_1,\V_2\subset \RR^4$,
they write $\V_1>\V_2$ if the set of difference vectors
$\V_1-\V_2$ is contained in $W_R$.   
They then show that there holds\footnote{This is the content of Lemma
3.2 in~\cite{BBS}. The lemma treats only the case of Bosons with 
compact localization
explicitly but, as the authors assert, easily extends to the
case of Fermions and of localization in space-like cones. 
}
\begin{equation} 
G^+(f)\phi = 
\begin{cases} (G(f)\Omega\times\phi)\Out& \text{ if } \V(f)>
\V(\phi)\\
(G(f)\Omega\times\phi)\In&\text{ if } \V(\phi)> \V(f). 
\end{cases}
\label{eqG+1Scatt'} 
\end{equation}

Let us first reformulate the condition on the velocity supports 
in \eqref{eqG+1Scatt'}. For a momentum vector $p\in V_+$ define the 
corresponding velocity 
$v(p):=\big(1,\frac{\bfp}{\omega(\bfp)}\big)$, and for two momentum
vectors $p,q$ define $p>_{W_R}q$ if and only if 
\begin{equation} \label{eqOrder'1}
v(p)-v(q)\in W_R. 
\end{equation}
Then $\V(f)>\V(\phi)$ if and only if $p>_{W_R}q$ for all
$p\in\supp \hat f$ and $q\in\spec \phi$. 
Since the support of the Fourier transform of $G^+$ is contained in
the positive mass shell
$H_m^+$, one may of course replace $\supp \hat f$ with $\supp \hat
f\cap H_m^+$ in the definition~\eqref{eqVf} of $\V(f)$, and the
conclusion~\eqref{eqG+1Scatt'} still holds. 
Now for an arbitrary wedge $W=\Lambda W_R+a$ let us {define} $p>_W q$
if and only if 
\begin{equation} \label{eqOrder'}
\Lambda^{-1}p>_{W_R}\Lambda^{-1} q.
\end{equation}
By covariance, and using $U(g)^{-1} G(f)U(g)=U(g)^{-1} G
U(g)(g^*f)$, where $g^*$ is the pull-back action of $g\in\Potild$, one 
readily verifies that the result~\eqref{eqG+1Scatt'} of Borchers et 
al.\ is equivalent with our Lemma~\ref{BBS} --- with the relation
$>_W$ instead of the relation $\succ_W$ defined in Eq.~\eqref{eqOrder}. 
It remains to show that these two relations coincide on the mass shell. 
\begin{Lem} \label{p>qW1}
On the mass shell, the relation $\succ_W$ coincides with the 
relation $>_W$ defined in  \eqref{eqOrder'1}, \eqref{eqOrder'}. 
\end{Lem}
\begin{Proof}
In a first step, we consider the standard wedge $W_R$.  
{}For a point $p$ on the mass shell, $v(p)$ is just
$p/p^0$. Hence for $p,q\in H_m^+$ the relation $p>_{W_R}q$
is equivalent with $\frac{p^1}{p^0}>\frac{q^1}{q^0}$, 
and hence with $p\succ_{W_R}q$ by Lemma~\ref{p>q}.  
Let now $W=\Lambda W_R+a$ be an arbitrary wedge. Recall that
$p>_Wq$ means by definition $\Lambda^{-1}p >_{W_R} \Lambda^{-1} q$,
which we have just shown to be equivalent with 
$\Lambda^{-1}p\in W_R + \RR \Lambda^{-1} q$, that is, 
with $p\in\Lambda W_R+\RR q$. But $\Lambda W_R$ is just the translated 
wedge $W-a$, hence $p>_W q$ is equivalent with $p\succ_Wq$. 
\end{Proof}
\providecommand{\bysame}{\leavevmode\hbox to3em{\hrulefill}\thinspace}
\providecommand{\MR}{\relax\ifhmode\unskip\space\fi MR }
\providecommand{\MRhref}[2]{%
  \href{http://www.ams.org/mathscinet-getitem?mr=#1}{#2}
}
\providecommand{\href}[2]{#2}

\end{document}